\documentclass[sigconf]{acmart}
\AtBeginDocument{%
  }
\usepackage{makecell}
\usepackage{graphicx}
\usepackage{booktabs}
\usepackage{enumitem}
\usepackage{diagbox} 
\usepackage{cases}
\usepackage{multirow}
\usepackage{multicol}
\usepackage{adjustbox}
\usepackage{amsfonts}

\usepackage{subfig}
\usepackage{xcolor}
\usepackage{listings}
\usepackage{hyperref}
\usepackage{threeparttable}  
\setcopyright{acmlicensed}
\copyrightyear{2018}
\acmYear{2018}
\acmDOI{XXXXXXX.XXXXXXX}
\acmConference[Conference acronym 'XX]{Make sure to enter the correct
  conference title from your rights confirmation email}{June 03--05,
  2018}{Woodstock, NY}
\acmISBN{978-1-4503-XXXX-X/2018/06}

\definecolor{codegreen}{rgb}{0,0.6,0}      
\definecolor{codegray}{rgb}{0.5,0.5,0.5}  
\definecolor{codepurple}{rgb}{0.58,0,0.82}
\definecolor{codeblue}{rgb}{0.25,0.5,0.75}
\definecolor{codeorange}{rgb}{0.8,0.4,0}   
\definecolor{codeback}{rgb}{0.95,0.95,0.92}
\begin{document}

\title{RideGym: A Standardized Interface for \\ Real-World Large-Scale Ride-Sharing System}



\author{Zijian Zhao}
\affiliation{%
  \institution{The Hong Kong University of Science and Technology}
  \city{Hong Kong}
  \country{China}}
\email{zzhaock@connect.ust.hk}

\author{Yulong Hu}
\affiliation{%
  \institution{The Hong Kong University of Science and Technology}
  \city{Hong Kong}
  \country{China}}
\email{yhucm@connect.ust.hk}

\author{Sen Li*}
\affiliation{%
  \institution{The Hong Kong University of Science and Technology}
  \city{Hong Kong}
  \country{China}}
\email{cesli@ust.hk}







\renewcommand{\shortauthors}{Zhao et al.}

\begin{abstract}
Ride-sharing has become an essential component of modern urban transportation and has attracted significant attention across computer science, transportation, and management science. While the field spans a broad range of problems, such as driver relocation, dynamic pricing, and vehicle charging or fueling dispatch, the core challenge remains order assignment and trip bundling, which directly affect urban traffic efficiency and carbon emissions. Despite its importance, existing simulation platforms are typically tailored to specific operational studies or tightly coupled to a particular dispatch algorithm, and rarely expose a standardized, learning-friendly interface. As a result, most researchers still build customized environments from scratch, raising serious concerns about reproducibility and fair comparison, and incurring substantial redundant effort. To address this gap, we present RideGym, the first open-source, standardized Gym-style interface tailored to Multi-Agent Reinforcement Learning (MARL)-based order dispatch in real-world ride-sharing systems. By fully decoupling the environment from the dispatch algorithm, RideGym enables diverse learning-based and model-based methods to be developed and compared under identical, fully specified conditions. It supports efficient, large-scale city-level simulations on real road networks, and offers flexible configurations for vehicle attributes (e.g., personalized speeds and capacities), order specifications (e.g., multiple passengers per order), and automatic shortest-path routing. We validate RideGym by reproducing several baselines, and demonstrate its high efficiency, with a one-hour simulation involving thousands of vehicles and tens of thousands of orders completed within one minute across all methods. Moreover, we reveal that the choice of exploration noise can significantly affect both the performance and the relative ranking of MARL solutions, an aspect often overlooked in prior work. Our code is available at \url{https://github.com/RS2002/RideGym}, and the Python package can be installed via \texttt{pip install ride-gym}.
\end{abstract}




\maketitle

\section{Introduction}

Ride-sharing services have fundamentally reshaped urban transportation over the past decade. Platforms such as Uber, Lyft, and DiDi provide on-demand mobility that offers travelers greater convenience, reduced waiting times, and improved accessibility compared to traditional taxis or public transit. For drivers, these platforms create flexible income opportunities, while for cities they influence traffic patterns, vehicle utilization rates, and overall carbon emissions. By enabling more efficient matching between supply and demand, ride-sharing has become an indispensable component of modern smart-city ecosystems, affecting daily commutes, economic productivity, and environmental sustainability at a massive scale \cite{jin2018ridesourcing,shaheen2019shared}.

Within ride-sharing systems, a variety of operational tasks have attracted substantial research attention, including driver relocation \cite{sun2022optimizing}, dynamic pricing \cite{ge2025marl}, and vehicle charging/fueling dispatch \cite{yang2026graph}. Among these tasks, order assignment and bundling form the core operational engine. They determine which available driver should serve each dynamically arriving passenger request and whether multiple orders can be combined into a single trip. As a result, they directly affect key performance metrics such as passenger waiting time, vehicle utilization, total vehicle miles traveled, traffic congestion, and greenhouse gas emissions. Effective order dispatch therefore lies at the heart of both platform profitability and broader societal impact \cite{alonso2017demand}.

However, ride-sharing order dispatch presents formidable challenges. From an operations research perspective, it is a highly dynamic matching problem under uncertainty: decisions must account for Origin-Destination (OD) relationships, the complex interactions between newly assigned orders and en-route orders (whose routes and capacities are already partially committed), and the stochastic nature of future demand and traffic conditions. Technically, the problem is exacerbated by an enormous joint state space, encompassing vehicle locations, passenger requests, road network states, and driver statuses, and an exponentially large action space, where each action may involve assigning individual orders, forming bundles, or choosing not to serve certain requests. These characteristics render exact solutions intractable at city scale and make real-time decision-making particularly demanding.

Despite these advances, progress in the field is held back less by a shortage of algorithms than by the absence of a common environment on which to develop and compare them. Several ride-sharing simulators do exist, yet none offers a standardized, learning-friendly testbed for order dispatch, and they broadly fall into two families, each missing a critical property. \emph{Operations-oriented} simulators \cite{chen2025hrsim, engelhardt2022fleetpy, kucharski2022simulating} faithfully model real road networks and high-capacity pooling, but are designed to evaluate operational strategies such as pricing, matching, and repositioning; they expose no standardized \texttt{reset}/\texttt{step} interface and are tightly coupled to a specific control logic, making it cumbersome to plug in and fairly compare (MA)RL algorithms. \emph{Learning-oriented} simulators, in contrast, are typically bundled with one particular algorithm behind a bespoke interface, and several abstract the city into a grid or hexagonal world rather than a real road network \cite{lin2018efficient}, limiting both fidelity and reusability. Consequently, most researchers still build custom simulators from scratch, leading to inconsistent experimental setups, unfair comparisons, and substantial redundant engineering effort, which ultimately undermines reproducibility and slows the community's progress. (A detailed review of existing works are deferred to Appendix~\ref{sec:related}.)

These observations point to a specific missing ingredient: a standardized, Gym-style interface \cite{brockman2016openai}. The \texttt{reset}/\texttt{step} abstraction that catalyzed progress in single-agent RL, and more recently in transportation MARL such as traffic-signal control, confers two properties that ride-sharing dispatch urgently needs. First, it decouples the environment from the decision algorithm: researchers can swap dispatch policies, whether model-based or (MA)RL, behind an identical interface, so that observed performance differences reflect the algorithms themselves rather than idiosyncrasies of independently re-implemented simulators. Second, it lowers the engineering barrier and enforces fully specified, reproducible conditions, allowing the community to accumulate directly comparable results instead of repeatedly rebuilding environments. Such standardization is arguably even more valuable, and harder to attain, in ride-sharing than in conventional single-agent settings: dispatch actions are coupled across agents (an order cannot be assigned to two vehicles simultaneously) and unfold on a real road network with dynamically committed en-route trips. This is precisely the gap RideGym is designed to fill.

To fill this gap, we present RideGym, the first open-source, standardized Gym-style interface \cite{brockman2016openai} tailored to (MA)RL-based order dispatch in real-world ride-sharing systems. By fully decoupling the environment from the dispatch algorithm, RideGym enables model-based and learning-based methods alike to be developed and compared under identical, fully specified conditions. It supports efficient, large-scale city-level simulations (e.g., Manhattan) with over one thousand vehicles operating on real road networks, and offers flexible vehicle configurations (e.g., personalized speeds and capacities), order specifications with multiple passengers per order, and automatic shortest-path routing with realistic travel-time estimation. We validate the environment by reproducing several classical model-based and competitive MARL-based baselines. Results demonstrate both the effectiveness of existing methods and the exceptional computational efficiency of our simulator, where one-hour simulations with thousands of vehicles and tens of thousands of orders complete within one minute across all tested methods. Furthermore, our benchmarking study reveals that the choice of exploration noise can significantly affect both the performance and the relative ranking of MARL-based solutions, an aspect often overlooked in prior work.

\section{Preliminary: Problem Formulation}
\label{sec:setup}

In our simulator, we consider a ride-sharing system comprising a large number of vehicles and orders (i.e., travel requests) arriving dynamically over time. Each vehicle has a fixed maximum capacity and an average moving speed. At each decision step, a vehicle may be assigned zero or more orders, subject to its capacity constraint, and subsequently optimizes its route by minimizing the total travel time across both en-route and newly assigned orders. For orders, passengers submit requests to the system at arbitrary times, and the system adds them to a pooling buffer for dispatch at the next decision step. Unconfirmed orders remain in the pool and await future decisions; however, passengers are impatient, and any order that remains unconfirmed beyond a waiting-time threshold will be withdrawn, incurring a potential loss for the platform.

We model each vehicle as an agent and formulate the problem as a Multi-Agent Markov Decision Process (MAMDP) \cite{littman1994markov}, denoted by \(<n,S,U,\mathcal{P},\mathrm{R},\gamma,O,T>\), where the components represent the number of agents (vehicles), joint state, joint action, joint state transition function, joint reward function, discount factor, joint observation, and time horizon, respectively. Although each agent corresponds to a vehicle, the agents collectively serve the platform's objective rather than individual drivers. Consequently, the system is fully centralized, and every agent has access to the complete global state, i.e., the joint observation \(O\) is equivalent to the joint state \(S\). The detailed formulation is as follows:

\noindent  \textbf{1) State \(S\):} At time step \(t\), the joint state consists of the individual vehicle states \(s^v_{i,t}\), the order-pooling state \(s^o_{t}\) (containing unassigned orders), and the current timestamp \(t\), expressed as:
\begin{equation}
\begin{aligned}
S_t = [s^o_{t},\, s^v_{1,t},\, s^v_{2,t},\, \ldots,\, s^v_{n,t},\, t].
\label{eq:state}
\end{aligned}
\end{equation}
The order-pooling state \(s^o_t\) is further defined as:
\begin{equation}
\begin{aligned}
s^o_{t} = [s^o_{1,t},\, s^o_{2,t},\, \ldots,\, s^o_{m_t,t}],
\label{eq:order}
\end{aligned}
\end{equation}
where \(m_t\) is the number of orders pending assignment at time \(t\). The state of vehicle \(i\), denoted \(s^v_{i,t}\), consists of:
\begin{equation}
\begin{aligned}
s^v_{i,t} = [l^p_{i,t},\, c_i^m,\, c_{i,t}^r,\, v_i,\, s^e_{i,t}],
\label{eq:vehicle}
\end{aligned}
\end{equation}
where \(l^p_{i,t}\) is the two-dimensional current position, \(c_i^m\) and \(c_{i,t}^r\) are the maximum and remaining capacities, \(v_i\) is the average speed, and \(s^e_{i,t}\) represents the en-route orders, given by:
\begin{equation}
\begin{aligned}
s^e_{i,t} = [s^e_{i,1,t},\, s^e_{i,2,t},\, \ldots,\, s^e_{i,k_{i,t},t}],
\label{eq:enroute}
\end{aligned}
\end{equation}
with \(k_{i,t}\) denoting the number of en-route orders for vehicle \(i\) at time \(t\). Each order \(o\), whether in \(s^o_t\) or \(s^e_{i,t}\), is represented by:
\begin{equation}
\begin{aligned}
o = [l^o,\, l^d,\, \tau^a,\, \tau^r,\,h],
\label{eq:single_order}
\end{aligned}
\end{equation}
where \(l^o\) and \(l^d\) are the two-dimensional origin and destination coordinates, \(\tau^a\) is the request time, \(\tau^r\) is the expected remaining travel time (set as 0 for orders in \(s^o_t\)), and \(h\) is the number of passengers for this order.

\noindent  \textbf{2) Action \(U\):} At time \(t\), the joint action \(U_t\) consists of individual actions:
\begin{equation}
\begin{aligned}
U_t = [u_{1,t},\, u_{2,t},\, \ldots,\, u_{n,t}] \in \{0,1\}^{n \times (m_t+1)},
\label{eq:action}
\end{aligned}
\end{equation}
where \(u_{i,t} \in \{0,1\}^{m_t+1}\). Specifically, \(u_{i,j,t}=1\) for \(j \leq m_t\) indicates that the \(j\)-th order in the pool is assigned to vehicle \(i\), while \(u_{i,m_t+1,t}=1\) indicates that vehicle \(i\) receives no order at this step.

Unlike standard MAMDPs where agents act independently, the joint action \(U_t\) not only has a time-varying dimensionality but must also satisfy the following constraints:
\begin{subequations} 
\begin{align}
 \sum_{i \in \mathcal{I}} u_{i,j,t} &\le 1, & \forall j &\in \mathcal{J}_t, \label{match_order} \\
\sum_{j \in \mathcal{J}_t \cup \{m_t+1\}} u_{i,j,t} &\ge 1, & \forall i &\in \mathcal{I}, \label{match_driver} \\
u_{i,m_t+1,t} \sum_{j \in \mathcal{J}_t} u_{i,j,t} &= 0, & \forall i &\in \mathcal{I}, \label{match_driver2} \\
\sum_{j \in \mathcal{J}_t} u_{i,j,t} \, h_{j,t} &\le c_{i,t}^r, & \forall i &\in \mathcal{I}, \label{capacity} \\
u_{i,j} &\in \{0,1\}, & \forall i &\in \mathcal{I},\ \forall j \in \mathcal{J}_t \cup \{m_t+1\},
\end{align}
\end{subequations}
where \(\mathcal{I} = \{1,\dots,n\}\) and \(\mathcal{J}_t = \{1,\dots,m_t\}\) are the sets of vehicle and order indices, respectively, and \(h_{j,t}\) denotes the passenger count of order \(j\) in the pool at time \(t\).

Eq.~\eqref{match_order} ensures that each order is assigned to at most one vehicle. Eq.~\eqref{match_driver}–\eqref{match_driver2} guarantee that each vehicle either takes no order or receives at least one, while Eq.~\eqref{capacity} enforces that the total number of assigned passengers does not exceed the remaining capacity.

\noindent \textbf{3) State Transition Function \(\mathcal{P}(\cdot|\cdot,\cdot)\):} The transition function \(\mathcal{P}(\cdot|S_t,U_t)\) specifies the probability distribution over next states. In our simulator, vehicle state transitions are deterministic: each vehicle moves along the planned shortest route. The only source of stochasticity is the order-pooling state \(s^o_{t+1}\), into which new orders arrive randomly. For simplicity and alignment with most existing work, we currently do not model complex transportation-network dynamics such as traffic congestion or accidents.

\noindent  \textbf{4) Reward Function \(\mathrm{R}(\cdot,\cdot)\):} Since the ride-sharing order dispatch task is fully cooperative, the global reward is the sum of individual vehicle rewards:
\begin{equation}
\begin{aligned}
\mathrm{R}(S_t,U_t) = \sum_{i \in \mathcal{I}} \mathrm{r}(s^v_{i,t}, u_{i,t}),
\label{eq:reward}
\end{aligned}
\end{equation}
where \(\mathrm{r}(\cdot,\cdot)\) is an individual reward function that can be customized according to user preferences over criteria such as customer waiting time, service rate, or detour time.

A default reward function, designed to reflect real-world ride-sharing market dynamics, is provided in our simulator:
\begin{equation}
\begin{aligned}
\mathrm{r}(s^v_{i,t}, u_{i,t}) = & \sum_{o \in \mathcal{O}(u_{i,t})} \Big( \beta_1 + \beta_2 \cdot \mathrm{Dis}(o[l^o], o[l^d]) \cdot o[h] \\
& - \beta_3 \cdot \mathrm{ExpectedTime}(o; s^v_{i,t}, \mathcal{O}(u_{i,t})) \Big) \\
& - \sum_{o \in s^e_{i,t}}\beta_4 \cdot \mathrm{ExtraTime}(o; s^v_{i,t}, \mathcal{O}(u_{i,t})),
\label{eq:individual_reward}
\end{aligned}
\end{equation}
where \(\beta_1\) to \(\beta_4\) are non-negative hyperparameters, \(\mathcal{O}(u_{i,t})\) is the set of orders assigned by action \(u_{i,t}\), and \(o[l^o], o[l^d], o[h]\) are the origin, destination, and passenger count as defined in Eq.~\eqref{eq:single_order}. The function \(\mathrm{Dis}(\cdot,\cdot)\) computes the distance between origin and destination; \(\mathrm{ExpectedTime}(\cdot;\cdot,\cdot)\) gives the expected travel time for the assigned order bundle under shortest-path routing (i.e., the time from assignment to drop-off minus the request time); and \(\mathrm{ExtraTime}(\cdot;\cdot,\cdot)\) measures the additional detour time incurred for all en-route orders due to route updates when inserting the new bundle.

This reward structure reflects real-world considerations: \(\beta_1 + \beta_2 \cdot \mathrm{Dis}(o[l^o], o[l^d]) \cdot o[h]\) represents platform revenue, proportional to trip distance and passenger count; \(\beta_3 \cdot \mathrm{ExpectedTime}(\cdot;\cdot,\cdot)\) and \(\beta_4 \cdot \mathrm{ExtraTime}(\cdot;\cdot,\cdot)\) capture passenger satisfaction, which may influence future platform retention. Typically, \(\beta_4 > \beta_3\), as passengers are more sensitive to uncertain and dynamically increasing travel times. Notably, \(\mathrm{ExtraTime}\) can be negative for some orders due to path optimization when new orders are inserted. Vehicles receiving no new orders at a given step receive zero reward, as defined in Eq.~\eqref{eq:individual_reward}.

\noindent  \textbf{5) Objective \(\mathrm{J}(\cdot)\):}
For a trajectory \(\mathcal{T} = \{S_1,U_1, S_2, U_2, \ldots, S_T, U_T\}\), the return is defined as:
\begin{equation}
\begin{aligned}
\mathrm{G}(\mathcal{T}) = \sum_{t=1}^T \gamma^{t-1}\mathrm{R}(S_i, U_i),
\label{eq:return}
\end{aligned}
\end{equation}
and the objective of a joint policy \(\Pi\) is:
\begin{equation}
\begin{aligned}
\mathrm{J}(\Pi) = \mathbb{E}_{\mathcal{T} \sim \pi} \big[ \mathrm{G}(\mathcal{T}) \big],
\label{eq:objective}
\end{aligned}
\end{equation}
with the goal of finding an optimal policy \(\Pi^*\) that maximizes \(\mathrm{J}(\Pi)\).

\section{RideGym: Library Construction}

\begin{figure}[htbp]
\centering 
\includegraphics[width=0.45\textwidth]{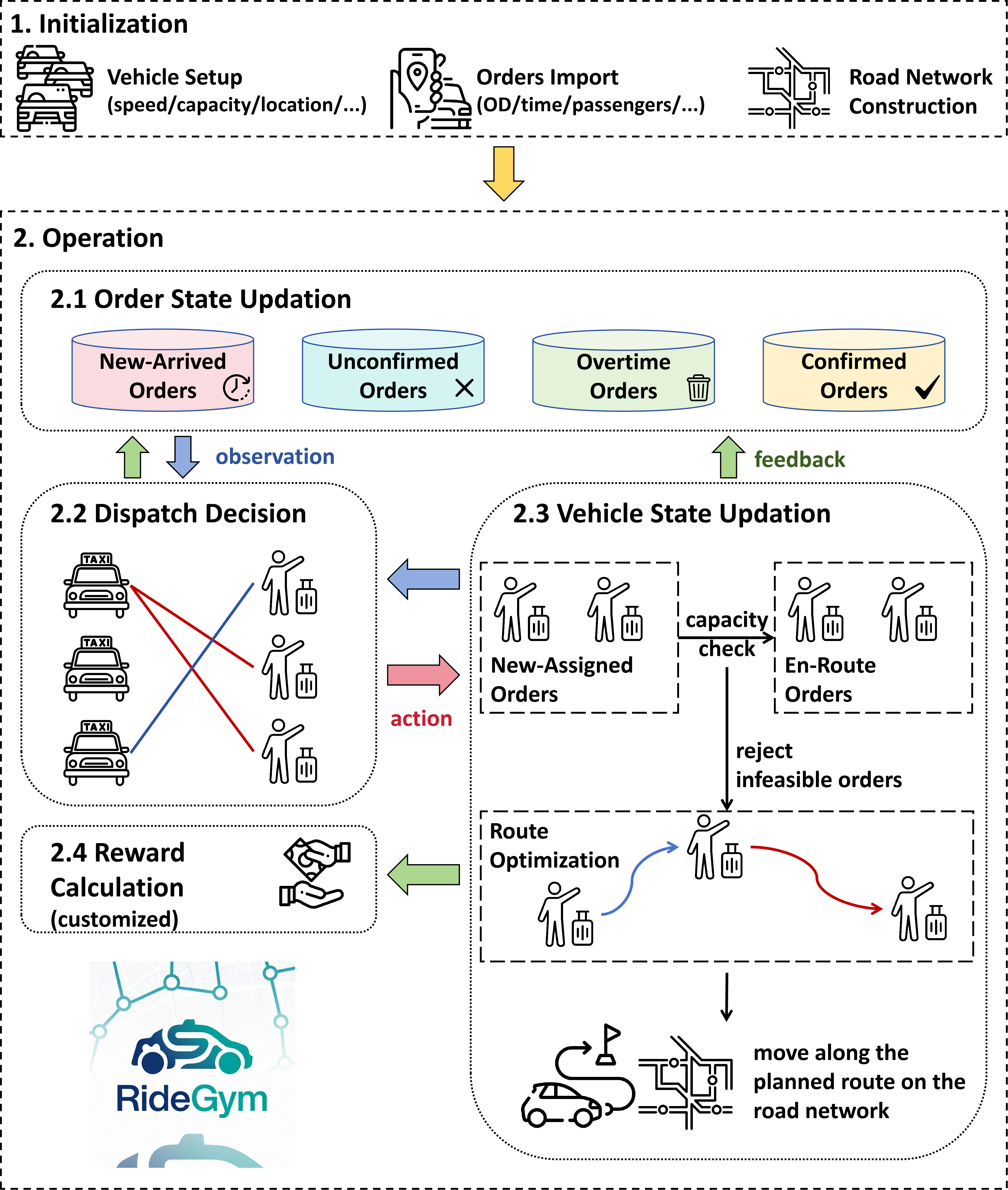}
\caption{Workflow of the proposed simulation framework.}
\label{fig:main}
\end{figure}


As shown in Fig.~\ref{fig:main}, the workflow of our simulator can be summarized as the following steps: \textbf{Step 1:} The simulator initializes the system state according to user-specified configurations, including vehicle attributes (e.g., personalized speed, capacity, and initial location), the order dataset (e.g., order origin-destination, arrival time, and passenger count), the operational region for road network construction, and the reward function. Unspecified parameters are set to random or default values. \textbf{Step 2.1:} At each time step, the order pool is updated: newly arrived orders are added, overdue unconfirmed orders are removed, and the status of en-route and completed orders is refreshed. \textbf{Step 2.2:} The user can then specify the order dispatch decisions through a standardized interface. \textbf{Step 2.3:} Based on the chosen actions, vehicle states are updated in the following order: conflicts between en-route and newly assigned orders are resolved, the shortest path for the updated bundle is re-optimized, and vehicles move along the road network following the planned routes. \textbf{Step 2.4:} The reward function computes the reward based on the intermediate outcomes from Step 2.3. Steps 2.1 through 2.4 repeat until the end of the time horizon, with the option for users to integrate training procedures into this iterative process. In this section, we provide a detailed description of the implementation of the core components and the utilization of our standard Application Programming Interface (API).

\subsection{Environment Implementation}
\label{sec:env-impl}

Our simulator is organized as a set of loosely coupled, interface-driven components, enabling independent replacement of each module without modifying the core simulation loop. This design is essential for rapidly constructing personalized scenarios in future research. The package comprises five core modules: (i) a set of lightweight domain \textbf{entities} (order and vehicle) that encapsulate the full lifecycle state of each request and vehicle; (ii) an \textbf{order generator} interface that supplies the demand stream for an episode; (iii) a \textbf{road network} interface that handles distance and shortest-path queries; (iv) a \textbf{route planner} interface that sequences each vehicle's pickups and drop-offs; and (v) a \textbf{reward function} interface that maps a per-step event log to the individual reward. These components are assembled by a central environment that owns all mutable state and enforces the strict event ordering illustrated in Fig.~\ref{fig:main}. Each interface is implemented as an abstract base class with a fast default implementation, allowing users to inject custom demand models, road backends, planners, or reward functions simply by passing an alternative object at construction time; the simulation loop remains agnostic to the concrete implementation used.

Concretely, the environment realizes the MAMDP formulated in Section~\ref{sec:setup} for a fleet of \(n\) vehicles \(\mathcal{I} = \{1, \dots, n\}\) over a horizon of \(T = \lceil H / \Delta t \rceil\) decision steps, where \(H\) is the total time horizon and \(\Delta t\) is the decision interval. At each step, it materializes the joint state \(S_t\) from Eq.~\eqref{eq:state}, receives the joint action \(U_t\), applies the deterministic vehicle transition alongside the stochastic arrival of new orders, and emits the reward \(\mathrm{R}(S_t, U_t)\). Below, we detail the two core components of our simulator: the precomputed road network and the precedence-aware route planner, and then outline the state-transition procedure.

\noindent \textbf{1) Precomputed Road Network:}
The road network is a performance-critical component: the distance query \(\mathrm{Dis}(\cdot, \cdot)\) is invoked on the order of millions of times per episode, as it underpins dispatch matching, route planning, and vehicle movement. Therefore, we aim to minimize the computational cost of this operation. In contrast to prior approaches \cite{hu2025bmg,wang2025demand} that issue per-query calls to external routing services such as OSRM \cite{luxen-vetter-2011}, our design avoids several inherent limitations: limited platform support (e.g., the official OSRM repository \footnote{\url{https://github.com/Project-OSRM/osrm-backend}} supports Linux but not Windows), high query latency, and the risk of packet loss or communication bottlenecks under parallel querying.

To address these issues, we adopt an ahead-of-time all-pairs shortest-path scheme. Given a real-world road network extracted from OpenStreetMap \cite{haklay2008openstreetmap}, we take its largest strongly connected component, relabel the nodes to contiguous indices \(\{1, \dots, K\}\), and precompute once the dense all-pairs shortest-path distance matrix \(\mathbf{D} \in \mathbb{R}^{K \times K}\) along with a predecessor matrix \(\mathbf{P} \in \mathbb{Z}^{K \times K}\), via \(K\) length-weighted Dijkstra passes:
\begin{equation}
\begin{aligned}
& \mathbf{D}_{ab} = \min_{\pi: \, a \rightsquigarrow b} \sum_{(u,w) \in \pi} \ell(u,w), \\
& \mathbf{P}_{ab} = \text{predecessor of } b \text{ on the shortest } a \rightsquigarrow b \text{ path},
\end{aligned}
\end{equation}
where \(\pi\) denotes a path in the road network, and \(\ell(u,w)\) is the physical length of directed edge \((u,w)\). Both matrices are cached to disk keyed by the node ordering, so that repeated runs load them in sub-second time, and any change to the graph automatically invalidates the cache. Any continuous query point, such as a vehicle position \(l^p_{i,t}\) or an order endpoint \(l^o, l^d\), is mapped to its nearest graph node via \(\mathrm{snap}(l) = \arg\min_k \| l - \chi_k \|_2^2\), where \(\chi_k\) are the node coordinates; this mapping is memoized, as query coordinates recur frequently. (The time to travel to the nearest node is accounted for through linear interpolation.) Consequently, the network distance reduces to two cached snap operations and a single matrix lookup:
\begin{equation}
\mathrm{Dis}(a,b) = \mathbf{D}_{\mathrm{snap}(a), \, \mathrm{snap}(b)},
\end{equation}
and a full node path is reconstructed in \(O(|\pi|)\) from \(\mathbf{P}\) without per-query Dijkstra. This design offers two key advantages over external routing services: it is faster (constant-time lookups instead of network round-trips and repeated search), and it is self-contained and system-independent, requiring no external service, daemon, or platform-specific binary. Travel time is computed as \(\mathrm{Dis}(a,b) / v_i\) minutes, decoupling distance topology from the vehicle speed model. For lightweight or purely abstract experiments, we additionally provide closed-form Euclidean and Manhattan backends behind the same interface, bypassing the graph entirely.

\noindent \textbf{2) Shortest-Path Route Planning under Precedence Constraints:}
When vehicle \(i\) is assigned the order bundle \(\mathcal{O}(u_{i,t})\), the environment must sequence the pickup and drop-off stops for both newly assigned and en-route orders \(s^e_{i,t}\) into a single coherent route. Let \(\mathcal{S}_{i,t}\) denote the resulting multiset of remaining stops, each either a pickup or a drop-off, subject to the ride-pooling precedence constraint that every order must be picked up before it is dropped off. This modeling choice enhances the fidelity of our simulator in reflecting real-world operations, in contrast to prior works \cite{hu2025bmg,zhao2025one} that prohibit vehicles from accepting new orders while en route to a pickup, a restriction often adopted for implementation simplicity.

Denoting by \(\sigma\) a visiting permutation of \(\mathcal{S}_{i,t}\) and by \(\mathrm{pos}_\sigma(\cdot)\) the position of a stop under \(\sigma\), the planner solves
\begin{equation}
\begin{aligned}
& \min_{\sigma} \Big[ \mathrm{Dis}(l^p_{i,t}, \sigma_1) + \sum_{k=1}^{|\mathcal{S}_{i,t}|-1} \mathrm{Dis}(\sigma_k, \sigma_{k+1}) \Big]
\quad \\
\text{s.t.} \quad & \mathrm{pos}_\sigma(\mathrm{pick}_o) < \mathrm{pos}_\sigma(\mathrm{drop}_o), \quad \forall o,
\label{eq:route}
\end{aligned}
\end{equation}
i.e., it minimizes the total route travel distance from the vehicle's current position \(l^p_{i,t}\), while respecting pickup-before-drop-off for every order. The default planner solves Eq.\eqref{eq:route} via a nearest-feasible-stop heuristic: starting from \(l^p_{i,t}\), it iteratively appends the nearest feasible stop, where a drop-off becomes feasible only after its corresponding pickup has been placed, thereby guaranteeing precedence by construction \cite{hurkens2004nearest}. This heuristic choice is adopted in place of an exact precedence-constrained Traveling Salesman Problem (TSP) solver because the planner resides on the hot path (invoked millions of times per episode) and the stop set size is bounded by the vehicle capacity \(c^m_i\). Consequently, the heuristic delivers near-optimal performance at a fraction of the cost, while the precedence constraint remains strictly enforced through the feasible-set restriction at each step. The same pass accumulates travel time for each leg, enabling the planner to return, at no additional cost, the predicted arrival times at every pickup and drop-off. As the planner is exposed through a generic interface, users may substitute an exact constrained-TSP solver while preserving the same precedence guarantee.

\noindent \textbf{3) State Transition:}
Given the joint action \(U_t\), the environment executes the transition \(\mathcal{P}(\cdot \mid S_t, U_t)\) in the strict order illustrated in Fig.~\ref{fig:main}: it (i) resolves conflicts to enforce constraint Eq.~\eqref{match_order} and Eq.~\eqref{capacity}, ensuring that no order is assigned to more than one vehicle and that assigned orders do not exceed the remaining capacity; (ii) admits each accepted bundle \(\mathcal{O}(u_{i,t})\) and re-solves Eq.\eqref{eq:route} for shortest-path route planning; and (iii) advances each vehicle \(i\) along its planned route by a distance budget \(v_i \Delta t\). On a real graph, a vehicle walks the reconstructed node path edge by edge and may halt part-way along an edge, with its continuous position \(l^p_{i,t}\) obtained by linear interpolation between the edge endpoints; a long edge is thus consumed across several steps, ensuring deadlock-free progress whenever \(v_i \Delta t > 0\). Arrivals trigger pickup and drop-off updates that advance each order through its lifecycle and decrement its expected remaining travel time \(\tau^r\). The vehicle transition is deterministic; the only stochasticity is the arrival of new requests into \(s^o_{t+1}\), while pending orders whose waiting time exceeds the threshold \(\theta\) are withdrawn before the next decision. (iv) Finally, the reward function maps the resulting per-vehicle events to the individual reward \(\mathrm{r}(s^v_{i,t}, u_{i,t})\).

\subsection{Standardized Interface}
\label{sec:api}
The environment exposes a decentralized, multi-agent, Gym-like API that is not dependent on the OpenAI Gym library \cite{brockman2016openai}. This section walks through a typical order of use, illustrated by the code a user would actually write in Fig. \ref{fig:env_api}.

\noindent \textbf{1) Preparing the Demand Dataset:}
The order stream is supplied through an order generator. Users may supply their own dataset (e.g., real historical trips) as a table with one row per order and the columns listed below, corresponding to the fields in Eq.~\eqref{eq:single_order} (\(l^o, l^d, \tau^a, h\); the remaining travel time \(\tau^r\) is initialized to \(0\) internally). A procedural generator (supporting uniform, Poisson, or peaked arrivals) is also provided for synthetic experiments.

\noindent \textbf{2) Building the road network.}
For real-map experiments, the road network is built once, offline, and cached to disk (the all-pairs matrices \(\mathbf{D}, \mathbf{P}\) of Section~\ref{sec:env-impl}); every subsequent run reloads the cache and answers each \(\mathrm{Dis}(\cdot,\cdot)\) query in \(O(1)\). Abstract experiments instead use the closed-form Euclidean or Manhattan backends.

\noindent \textbf{3) Initializing the Environment:}
The environment is constructed from the components above together with the fleet configuration. Per-vehicle capacities \(c^m_i\) and speeds \(v_i\) may be provided as length-\(n\) lists for a heterogeneous fleet, or as scalars for a homogeneous one; the planner and reward function \(\mathrm{r}(\cdot,\cdot)\) are injected here and may be swapped freely.

\noindent \textbf{4) The \texttt{reset} and \texttt{step} Interface:}
Two methods drive the entire simulation:
\begin{align*}
(O_1,\, \mathrm{info}) &= \texttt{env.reset}(\mathrm{seed}),\\
(O_{t+1},\, \{\mathrm{r}(s^v_{i,t},u_{i,t})\}_{i\in\mathcal{I}},\,
\mathrm{done},\, \mathrm{info}) &= \texttt{env.step}(U_t).
\end{align*}
Both the observation and the reward are dictionaries keyed by vehicle index \(i \in \mathcal{I}\). Since the system is fully centralized, the returned observation coincides with the joint state, \(O_t \equiv S_t\) (Eq.~\eqref{eq:state}): for each vehicle, it contains the vehicle state \(s^v_{i,t}\) (Eq.~\eqref{eq:vehicle}) and the shared order-pool state \(s^o_t\) (Eq.~\eqref{eq:order}). The input \(U_t\) is the joint action of Eq.~\eqref{eq:action}, supplied as a dictionary of per-vehicle actions \(u_{i,t}\); each \(u_{i,t}\) is given compactly as the set of pool indices the vehicle bids on, and an empty set encodes the no-order action \(u_{i,m_t+1,t} = 1\). A single \texttt{step} advances the simulation by \(\Delta t\) and automatically enforces the feasibility constraints \eqref{match_order}-\eqref{capacity}.

\noindent \textbf{5) A Minimal Control Loop.}
Any policy that maps a vehicle state \(s^v_{i,t}\) and the order pool \(s^o_t\) to the action schema above can be rolled out as Fig. \ref{fig:env_api} (4).
A training procedure may be interleaved directly between successive \texttt{step} calls, matching the iterative workflow of Fig.~\ref{fig:main} and optimizing the objective \(\mathrm{J}(\Pi)\) of Eq.~\eqref{eq:objective}.

\begin{figure*}[htbp]
\centering
\begin{minipage}{\textwidth}
\small
\noindent \textbf{(1) Preparing the Demand Dataset}
\begin{lstlisting}[language=Python,basicstyle=\small\ttfamily]
from ride_gym.order_generator import DataFrameOrderGenerator
# Required columns:  origin_x, origin_y  -> l^o
#                    dest_x,   dest_y    -> l^d
#                    request_time        -> tau^a (minutes from episode start)
#                    num_passengers      -> h
order_gen = DataFrameOrderGenerator(dataframe=my_orders_df)
\end{lstlisting}

\vspace{0.3cm}
\noindent \textbf{(2) Building the Road Network}
\begin{lstlisting}[language=Python,basicstyle=\small\ttfamily]
from ride_gym.osmnx_network import OSMnxNetwork
# One-off offline build for a region, then reused via the on-disk cache:
#   python -m ride_gym.build_network --place "Manhattan, NY" --out data/manhattan.gpickle
network = OSMnxNetwork(graph_path="data/manhattan.gpickle")
\end{lstlisting}

\vspace{0.3cm}
\noindent \textbf{(3) Initializing the Environment}
\begin{lstlisting}[language=Python,basicstyle=\small\ttfamily]
from ride_gym.env import RidePoolEnv
env = RidePoolEnv(
    num_drivers=1000,                # n
    driver_capacities=[3, 4, ...],   # per-vehicle c^m_i (or driver_capacity=4)
    driver_speeds=[1.0, 0.9, ...],   # per-vehicle v_i   (or None -> network speed)
    dt=1.0, horizon=60.0,            # decision interval Delta t, horizon H
    order_timeout=3.0,               # impatience threshold
    order_generator=order_gen,
    road_network=network,
)
\end{lstlisting}

\vspace{0.3cm}
\noindent \textbf{(4) Minimal Control Loop}
\begin{lstlisting}[language=Python,basicstyle=\small\ttfamily]
obs, info = env.reset(seed=0)                # initial joint observation O_1 == S_1
done = False
while not done:
    U = {}                                   # joint action {i: u_{i,t}}
    for i, s_v in obs.items():
        pool = s_v["pending_orders"]         # shared order-pool state s^o_t
        u_i = my_policy(s_v, pool)           # user policy -> list of bid order ids
        U[i] = {"orders": u_i}               # empty list encodes the no-order action
    obs, rewards, dones, info = env.step(U)  # rewards[i] = r(s^v_{i,t}, u_{i,t})
    done = dones["__all__"]                  # True once step t reaches horizon T
\end{lstlisting}

\vspace{0.3cm}
\noindent \textbf{(5) Visualization}
\begin{lstlisting}[language=Python,basicstyle=\small\ttfamily]
from ride_gym.visualize import render_frame, TrajectoryRecorder, render_animation

# (a) Single frame -> static image (PNG or PDF).
env.render(mode="human", save_path="frame.pdf")

# (b) Whole episode -> animation. Snapshot each step, then export to GIF/MP4.
rec = TrajectoryRecorder()
obs, _ = env.reset(seed=0)
done = False
while not done:
    obs, rewards, dones, info = env.step(my_policy(obs))
    rec.snapshot(env)                 # lightweight per-step capture
    done = dones["__all__"]
render_animation(rec, out_path="episode.gif")   # or "episode.mp4"
\end{lstlisting}
\end{minipage}
\caption{Typical Python usage workflow.}
\label{fig:env_api}
\end{figure*}

\subsection{Visualization}
\label{sec:visual}
To make simulation dynamics interpretable, our environment ships with a built-in visualization toolkit. It can render either a single frame (a static snapshot of the system state at one decision step) or an entire episode as an animation, and exports to standard formats: a static image (PNG/PDF) for a frame, and an animated GIF or MP4 for a rollout. Rendering reads only the public state already exposed by the environment, so it is fully decoupled from the simulation loop and adds no overhead when disabled. A typical use is a one-line call after (or during) a rollout, shown in Fig. \ref{fig:env_api} (5). A detailed example is provided at Appendix \ref{sec:example}. Beyond the spatial view, our toolkit also provides a set of aggregate, metric-oriented visualizations, including demand and service-rate heatmaps, supply-demand gap maps, system-load time series, and passenger waiting-time distributions, for quantitative analysis of a scenario or a policy. As these are auxiliary to the core environment, we omit their details here and refer the interested reader to our repository for the full set of tools and usage examples.

\section{Benchmarking Experiment}
\subsection{Benchmark Approaches}

To validate the effectiveness of our proposed simulator, we evaluate a set of benchmark methods in our environment, covering both classical model-based approaches and recent MARL-based solutions. For model-based methods, we select widely used strategies from online ride-hailing platforms such as Didi \cite{chen2023didi}, including Random dispatch, Greedy \cite{kalyanasundaram1993online}, Kuhn–Munkres (KM) \cite{kuhn1955hungarian}, and Gale–Shapley (GS) \cite{gale1962college, yue2024end}. For MARL-based solutions, we choose REDA \cite{holder2025multi}, BMG-Q \cite{hu2025bmg}, MF-DDQN \cite{li2019efficient}, Assignment-Net \cite{zhao2025impacts, zhao2026discriminatory}, and CV-Net \cite{tang2019deep}, covering a range of innovations in both algorithmic design and network architecture. Detailed introductions to these methods are provided in Appendix~\ref{sec:benchmark}. Specifically, we emphasis the Random is not meaningless since the matching is happened in a given radius, so it can be viewed as another type of Greedy.

For all MARL-based methods, we introduce a dummy order that represents the option of taking no order at the current step. This design allows agents to actively reject low-value orders—for instance, when an order is not aligned with the en-route trajectories of nearby vehicles and would therefore negatively impact the system. This mechanism is not implemented in the original papers of these methods.

\subsection{Experiment Configurations}

The experiments are conducted using the public ride-hailing dataset from Manhattan, New York, provided by the TLC Trip Record Data \cite{TLCData}. We set the fleet size to 1,000 vehicles, each with a capacity of 4 passengers and an average speed of 35 km/h, reflecting typical urban driving conditions. Since the TLC data provides only coarse-grained zone-level OD information, we assign each OD to the central coordinate of its corresponding zone, with a random perturbation drawn uniformly from a circle of radius 0.5 km, to avoid unrealistic concentration of orders at identical points. The passenger count per order is sampled uniformly between 1 and 4, capturing the heterogeneity of real-world ride-sharing demand, which is an aspect often overlooked in prior work. Specifically, our simulator also supports a mixture of pooling and non-pooling orders, reflecting the practical scenario where passengers may opt out of ride-sharing by selecting a passenger count of 4 and paying a higher fare. The maximum waiting time for order confirmation is set to 3 minutes; any order not assigned to a vehicle within this window is considered canceled by the passenger. Each episode spans a 60-minute horizon with a decision interval of 1 minute, which is a common setup applied by many papers \cite{hu2025bmg,zhao2025impacts,al2019deeppool,enders2023hybrid,hoppe2024global}.

For MARL-based methods, we adopt the Adam optimizer with a learning rate of $5 \times 10^{-4}$. The hyper-parameters $gamma$ and $\beta_1$t o $\beta_4$ are set as $0.99,1.0,0.01,0.04,0.08$. The replay buffer size is set to 6,000 (equivalent to 100 episodes), with a batch size of 8, where each sample corresponds to the joint transition of all agents at a single time step. The exploration strategy is detailed in the experiment results section. Each method is trained for 500 episodes. We use data from 8:00 to 20:00 between April 6 and April 12, 2026, as the training set, April 13, 2026, as the validation set (optional, for tracking training progress), and April 14, 2026, as the testing set. Specifically, we report results on an off-peak scenario (10:00–11:00, with 6,863 orders) and an on-peak scenario (18:00–19:00, with 11,219 orders) from April 14, 2026. All data are sourced from the High Volume FHV Trips Data within the TLC Trip Record Data. All training and evaluation are conducted on a workstation running Windows 11, equipped with an Intel(R) Core(TM) i7-14700KF processor and an NVIDIA RTX 4080 graphics card.

\subsection{Evaluation Results}

\begin{table*}[t!]
\centering
\caption{Benchmark performance under off-peak and on-peak settings. $\uparrow$ / $\downarrow$ denote whether higher or lower is better. All time metrics are in minutes. Best and second-best results per column are marked in \textbf{bold} and \underline{underline}, respectively. Metric definitions are given in Appendix~\ref{sec:metric}. All results are averaged over three independent runs.}
\label{tab:benchmark}
\setlength{\tabcolsep}{5.5pt}
\renewcommand{\arraystretch}{1.15}
\begin{adjustbox}{width=\textwidth}
\begin{tabular}{l ccccc ccccc}
\toprule
\multirow{2}{*}{\textbf{Method}}
 & \multicolumn{5}{c}{\textbf{Off-Peak (10:00-11:00 with 6,863 orders)}}
 & \multicolumn{5}{c}{\textbf{On-Peak (18:00-19:00 with 11,219 orders)}} \\
\cmidrule(lr){2-6} \cmidrule(lr){7-11}
 & Service & Completion & Wait & Detour & Simulation
 & Service & Completion & Wait & Detour  & Simulation\\
 & Rate (\%) $\uparrow$ & Rate (\%) $\uparrow$ & Time $\downarrow$ & Time $\downarrow$ & Time $\downarrow$
 & Rate (\%) $\uparrow$ & Rate (\%) $\uparrow$ & Time $\downarrow$ & Time $\downarrow$  & Time $\downarrow$\\
\midrule
\multicolumn{11}{l}{\emph{\textbf{Model-based methods}}} \\
Random          & 91.21±0.19  & 77.63±0.20 & 1.84±0.02  & 1.66±0.01 & 0.13±0.05 & 68.87±0.58 & 55.75±0.61 & 2.11±0.02 & 2.61±0.03 & 0.21±0.06 \\
Greedy          & 93.98±0.20 & 80.94±0.28 & 1.50±0.03 & 1.61±0.04 & 0.13±0.05 & 78.41±0.29 & 65.12±0.28 & 2.22±0.04 & 2.27±0.06 & 0.18±0.11 \\
KM              & 93.89±0.34 & 80.64±0.31 & 1.44±0.02 & 1.71±0.05 & 0.13±0.06 & 74.13±0.39 & 60.37±0.34 & 1.87±0.01 & 2.38±0.04 & 0.18±0.09 \\
GS              & 93.37±0.58 & 79.98±0.56  & 1.49±0.02 & 1.69±0.03 & 0.13±0.01 & 63.37±0.23 & 50.58±0.19 & 2.01±0.01 & 2.45±0.04 & 0.22±0.10 \\
\midrule
\multicolumn{11}{l}{\emph{\textbf{MARL-based methods (INF noise)}}} \\
REDA           & 91.04±0.38 & 79.24±0.27 & \textbf{1.14±0.01} & 1.13±0.01 & \textbf{0.08±0.02} & 74.97±0.55 & 61.79±0.41 & \textbf{1.28±0.01} & 2.33±0.04 & \underline{0.10±0.05} \\
BMG-Q           & 94.40±0.28 & 82.11±0.34 & 1.35±0.00 & 1.21±0.02 & 0.44±0.21 &  76.22±0.24 & 62.68±0.09 & \underline{1.40±0.00} & 2.39±0.05 & 0.45±0.03 \\
MF-DDQN         &  83.64±0.83 & 71.81±0.80 & \underline{1.15±0.01} & 1.29±0.01 & 0.45±0.013 & 71.58±0.36 & 58.86±0.27 & \textbf{1.28±0.00} & 2.47±0.01 & 0.50±0.16 \\
Assignment-Net  & 75.39±0.15 & 65.69±0.22 & 1.44±0.03 & \textbf{0.19±0.01} &  \textbf{0.08±0.03} & 73.69±0.29 & 64.76±0.43 & 1.42±0.00 & \textbf{0.22±0.01} & \underline{0.10±0.02} \\
CV-Net          &  87.88±0.29 & 76.00±0.40 & 1.29±0.01 & 1.23±0.03 & \textbf{0.08±0.03} &  75.12±0.66 & 61.68±0.70 & 1.48±0.02 & 2.19±0.01  & \underline{0.10±0.02} \\
\midrule
\multicolumn{11}{l}{\emph{\textbf{MARL-based methods (STD noise)}}} \\
REDA           & 96.50±0.25 & \textbf{85.68±0.18} & 1.40±0.02 & 0.69±0.02 & \textbf{0.08±0.02} & 78.96±0.41 & 64.70±0.48 & 1.58±0.02 & 2.20±0.01 & \textbf{0.09±0.04} \\
BMG-Q           & \underline{97.26±0.09} & 84.64±0.14 & 1.48±0.01 & 1.10±0.01 & 0.44±0.12 & 78.71±0.25 & 64.90±0.17 & 1.50±0.01 & 2.20±0.03 & 0.45±0.04 \\
MF-DDQN         & \textbf{97.49±0.11} & \underline{85.56±0.07} & 1.56±0.02 & 0.80±0.03 & 0.44±0.18 & \underline{80.16±0.09} & \underline{66.15±0.07} & 1.61±0.01 & 2.18±0.01 & 0.50±0.17 \\
Assignment-Net  & 95.83±0.09 & \textbf{85.68±0.14} & 1.44±0.01 & \underline{0.34±0.02}  & \textbf{0.08±0.02} & \textbf{87.18±0.28} & \textbf{77.30±0.31} & 1.80±0.01 & \underline{0.53±0.01} & \underline{0.10±0.03} \\
CV-Net          &  96.23±0.11 & 83.74±0.05 & 1.50±0.02 & 0.74±0.02 & \underline{0.09±0.03} & 79.42±0.40 & 65.37±0.40 & 1.69±0.02 & 2.16±0.05 & 0.27±0.03 \\
\bottomrule
\end{tabular}
\end{adjustbox}
\end{table*}

In this paper, we identify a critical factor that hinders the reproducibility of MARL-based ride-sharing methods: the design of exploration noise. Unlike standard MDP settings, actions in ride-sharing are not independent across agents, as an order cannot be assigned to multiple vehicles simultaneously. Consequently, common exploration techniques such as $\epsilon$-greedy or Boltzmann exploration are not directly applicable. Current methods typically adopt a paradigm in which noise is first added to the Q-values of each vehicle-order pair, followed by bipartite matching to maximize the global Q-value. Unfortunately, many prior works do not clearly specify this implementation detail. To systematically investigate its impact, we reproduce the \textbf{INF noise} from \cite{hu2025bmg}, where the Q-value of each vehicle-order pair is set to $+\infty$ with probability $\epsilon$ (the exploration rate, which decays from 1 to 0 over time), thereby forcing that pair to be selected. We further propose a simple yet effective \textbf{STD noise}, which adds a zero-mean Gaussian perturbation scaled by the current Q-value volatility:
\begin{equation}
\begin{aligned}
\zeta \sim \mathcal{N}(0, (\epsilon \sigma)^2),
\label{eq:std}
\end{aligned}
\end{equation}
where $\sigma$ is the standard deviation of the Q-values across all vehicle-order pairs.

The experimental results in Table~\ref{tab:benchmark} reveal two striking insights. First, STD noise consistently outperforms INF noise across all MARL baselines in terms of service rate and completion rate. While serving more orders inevitably leads to slightly higher wait and detour times (within 0.5 minutes), the substantial gains in service rate (ranging from 2\% to 20\%) far outweigh these marginal increases. We attribute this advantage to the adaptive scaling of STD noise, which aligns perturbation magnitude with the underlying Q-value distribution. In contrast, INF noise often induces conflicting simultaneous selections (e.g., multiple vehicles forced to the same order) that are later pruned by the matching step, resulting in insufficient and biased exploration. Second, and more critically, the choice of noise fundamentally alters the relative ranking of algorithms. For example, under INF noise, Assignment-Net and MF-DDQN underperform relative to other MARL baselines; however, under STD noise, they emerge as the strongest contenders. This suggests that conclusions drawn under arbitrary noise configurations may not be universally valid.

A closer examination of Table~\ref{tab:benchmark} further disentangles algorithmic strengths. Since STD noise yields consistently better overall performance, we consider it a more reliable setting for assessing converged performance and use it as the basis for our analysis. During off-peak periods, MF-DDQN and BMG-Q achieve marginally higher service rates than others, suggesting that neighborhood information aggregation via mean-field or graph attention is beneficial when demand is sparse and competition among vehicles is low. However, during on-peak periods, their advantages become less pronounced, likely because the abundance of orders reduces the risk of conflicting competition, thereby limiting the added value of neighbor-aware aggregation. Moreover, we observe that the per-episode simulation time for MF-DDQN and BMG-Q (0.44–0.50 min) is approximately five times higher than that of Assignment-Net and CV-Net (0.08–0.10 min). Although all methods comfortably complete a one-hour simulation within one minute, this overhead accumulates substantially during training: over 500 episodes, the extra cost amounts to over three hours of wall-clock time. In practice, when extensive hyper-parameter tuning is required, this computational gap becomes a non-negligible factor, and researchers must carefully balance the marginal performance gains of neighborhood aggregation against its significant training overhead.

Furthermore, we observe that network architecture design is also a critical factor in ride-sharing performance. Assignment-Net and CV-Net, with their carefully engineered structures, show superior detour performance, especially in on-peak scenarios. Specifically, Assignment-Net achieves exceptionally low detour times (0.34 min off-peak and 0.53 min on-peak), which are roughly two to four times lower than other methods (0.69–1.10 min off-peak and 2.16–2.20 min on-peak). We posit that this dramatic improvement stems from Assignment-Net's LSTM module, which explicitly preserves the sequential order of en-route pickups and drop-offs. In contrast, MLP- or GAT-based methods typically fuse sequential stop features into a fixed-length vector, discarding temporal order. This effect is amplified during peak hours, when the average number of en-route orders per vehicle increases, making order-preserving representations critical for efficient pooling.

Finally, the high computational efficiency of our simulator is validated by the fact that every benchmark method completes a full episode in under one minute across both scenarios, confirming that the environment itself will not become a bottleneck in large-scale MARL training or hyper-parameter search.

\section{Conclusion}


In this paper, we introduce RideGym, the first open-source and standardized Gym-style interface tailored to (MA)RL-based order dispatch in large-scale real-world ride-sharing systems. By fully decoupling the environment from the dispatch algorithm, RideGym provides a fair and efficient benchmarking platform for evaluating diverse methods under consistent, fully specified conditions, while also offering a rich set of modular APIs that serve as a flexible foundation for customized research, significantly reducing the need to build complex simulation systems from scratch. The simulator supports a wide range of configurable features, including personalized vehicle speeds, capacities, and multiple passengers per order, enabling realistic and configurable emulation of ride-sharing operations. We validate RideGym through extensive experiments with classical and competitive baselines, demonstrating its high computational efficiency and fast simulation speed. Notably, during the reproduction process, we uncover that the type of exploration noise can substantially affect both the absolute performance and the relative ranking of MARL-based methods—an important yet often overlooked factor in prior work. We believe RideGym will foster improved reproducibility, fairer comparisons, and reduced redundant engineering efforts in the intelligent transportation research community. More discussions are left at Appendix~\ref{sec:discussion}.

\bibliographystyle{ACM-Reference-Format}
\bibliography{sample-base}

\appendix

\section{Literature Reviews}
\label{sec:related}

\subsection{Model-Based Order Dispatch}
Order dispatch has traditionally been cast as a combinatorial optimization problem and solved with model-based methods. Early formulations assumed that requests were fully known ahead of time and matched trips to vehicles by solving linear assignment programs \cite{yan2011model, cordeau2007dial}. An influential line of work is the Request-Trip-Vehicle (RTV) framework of \citet{alonso2017demand}, which enumerates feasible order bundles, links them to compatible vehicles with cost-weighted edges, and obtains a cost-minimizing assignment by solving a bipartite matching problem, together with a demand-driven vehicle rebalancing step. Because constructing and solving the RTV program is expensive at scale, subsequent studies traded optimality for speed, e.g., restricting each vehicle to at most one new request per epoch and relying on implicit bundling of en-route and incoming orders \cite{simonetto2019real}. To counter the myopia of one-shot matching, later methods incorporated future information like demand forecasts appended to the assignment graph \cite{alonso2017predictive}, rolling-horizon and model-predictive control for joint relocation and dispatch \cite{riley2021real}, and stochastic-programming formulations under demand uncertainty. 

\subsection{Reinforcement Learning for Order Dispatch}
The complexity of modeling stochastic mobility systems has motivated model-free reinforcement learning (RL), which learns dispatch policies directly from interaction and implicitly captures uncertainty and long-horizon effects. \citet{xu2018large} first scaled RL to ride-hailing by learning a per-vehicle value function and recovering the assignment through global bipartite matching on Q-value-weighted edges; many works inherit this ``learn-a-value-then-match'' paradigm \cite{qin2020ride, hu2025bmg}. Owing to the high-dimensional joint state-action space, most methods adopt a MARL formulation, commonly grouped into Decentralized Training with Decentralized Execution (DTDE), Centralized Training with Decentralized Execution (CTDE), and Centralized Training with Centralized Execution (CTCE) \cite{jin2025comprehensive}. DTDE methods treat every vehicle as an independent learner \cite{al2019deeppool}, which is simple but suffers from non-stationarity and weak coordination; neighbor-aware encoders such as graph attention \cite{hu2025bmg} and mean-field approximations \cite{li2019efficient} partially alleviate this. CTDE and CTCE methods instead pursue stronger cooperation via centralized critics, global rewards, and value decomposition, adapting architectures such as MASAC \cite{enders2023hybrid}, COMA \cite{hoppe2024global}, and QMIX \cite{hao2022hierarchical} to the dispatch setting.

\subsection{Simulation Platforms}
Unlike domains such as traffic signal control or combinatorial optimization, where shared environments and benchmarks have accelerated progress (e.g., Gym-style RL interfaces \cite{brockman2016openai}, microscopic traffic simulators \cite{lopez2018microscopic}, and unified routing benchmarks such as RL4CO \cite{berto2025rl4co}), ride-sharing order dispatch still lacks a comparably standardized evaluation substrate. This is not for want of simulators: a number of open-source platforms exist, but each misses at least one property required to serve as a standardized (MA)RL dispatch benchmark, as summarized in Table~\ref{tab:sim_compare}.

Broadly, these platforms fall into two families. \emph{Operations-oriented} simulators are built to evaluate system-level strategies such as pricing, matching, and repositioning. HRSim \cite{chen2025hrsim} models high-capacity pooling on real road networks with a focus on pricing and emission studies; FleetPy \cite{engelhardt2022fleetpy} provides a mature modular pipeline with heterogeneous fleets and ILP-based pooling; MaaSSim \cite{kucharski2022simulating} targets two-sided market and behavioral dynamics; and the dispatcher of \citet{laupichler2026advancing} pushes raw scalability to tens of thousands of vehicles. These platforms are faithful and mature, but they expose no standardized \texttt{reset}/\texttt{step} interface and are tightly coupled to a specific control pipeline, so plugging in and fairly comparing (MA)RL policies requires substantial re-engineering. \emph{Learning-oriented} simulators, in contrast, are usually released as a byproduct of a particular algorithm: MOVI \cite{oda2018movi} and the widely used DiDi environment \cite{lin2018efficient} are each tied to a specific method behind a bespoke interface, and the latter further abstracts the city into a grid/hex world rather than a real road network, limiting fidelity and reusability. The closest platform in spirit is the multi-functional simulator of \citet{feng2024transportation}, which exposes RL ``portals'' for ride-sourcing operations, yet it offers bespoke interfaces rather than a standardized, algorithm-agnostic Gym API and targets single-occupancy ride-sourcing rather than order bundling. Finally, Grab has reported a Gym-style dispatch environment \cite{tan2025dispatchgym}, but it remains closed-source, with no code or interface released more than a year after the initial report.

\emph{These observations expose a precise gap: the field has many dispatch algorithms and several faithful simulators, yet none offers a standardized, reusable, algorithm-agnostic environment on which decentralized and centralized methods can be trained and compared under identical conditions. RideGym fills exactly this gap. It (i) supports both abstract (Euclidean/Manhattan) and real OpenStreetMap-based road networks with precomputed shortest paths for fast, system-independent distance queries; (ii) models heterogeneous fleets (per-vehicle speeds and capacities), multiple passengers per order, user-supplied demand datasets, and configurable rewards; and (iii) exposes a standardized, algorithm-agnostic, Gym-like \texttt{reset}/\texttt{step} interface, so that model-based and (MA)RL methods alike can be benchmarked reproducibly. As Table~\ref{tab:sim_compare} shows, RideGym is the only environment that satisfies all of these requirements simultaneously.}

\begin{table*}[t]
\centering
\caption{Comparison of open-source ride-sharing / ride-hailing simulators against the requirements for a standardized (MA)RL order-dispatch benchmark. \checkmark: supported; $\times$: not supported; $\sim$: partial.}
\label{tab:sim_compare}
\setlength{\tabcolsep}{6pt}
\renewcommand{\arraystretch}{1.15}
\begin{tabular}{l l cccccc}
\toprule
\textbf{Simulator} & \textbf{Primary purpose}
 & \makecell{Real road\\network}
 & \makecell{Hetero.\\fleet}
 & \makecell{Ride\\pooling}
 & \makecell{Multi-pax\\per order}
 & \makecell{Gym-style\\API}
 & \makecell{Algorithm\\-agnostic} \\
\midrule
DiDi env.\ \cite{lin2018efficient}     & Fleet rebalancing        & $\times$ & $\times$ & $\times$ & $\times$ & $\times$ & \textcolor{black}{$\times$} \\
MOVI \cite{oda2018movi}                & Dispatch \& repositioning & \textcolor{black}{\checkmark} & $\times$ & $\times$ & $\times$  & $\times$ & $\times$ \\
MaaSSim \cite{kucharski2022simulating} & Two-sided market         & \checkmark & \textcolor{black}{$\sim$} & \textcolor{black}{\checkmark} & \textcolor{black}{$\times$}   & $\times$ & $\times$ \\
FleetPy \cite{engelhardt2022fleetpy}   & MoD operations           & \checkmark & \checkmark & \checkmark & \textcolor{black}{$\times$}  & \textcolor{black}{$\sim$} & $\times$ \\
\citet{laupichler2026advancing}        & Large-scale dispatch     & \checkmark & \textcolor{black}{$\times$} & \checkmark & \textcolor{black}{$\times$}   & $\times$ & $\times$ \\
HRSim \cite{chen2025hrsim}             & Ride-sharing ops (matching, pricing) & \checkmark & \textcolor{black}{$\sim$} & \checkmark & \textcolor{black}{$\times$}   & $\times$ & $\times$ \\
\citet{feng2024transportation}         & Ride-sourcing RL testbed & \checkmark & \textcolor{black}{$\times$} & $\times$ & $\times$   & $\times$ & $\sim$ \\
Triple-BERT \cite{zhao2026triplebert} & RL order dispatch & \checkmark & $\times$ & \checkmark & $\times$ & $\times$ & $\times$ \\
OSPO \cite{zhao2025one} & MARL order dispatch & \checkmark & $\times$ & \checkmark & $\times$ & $\times$ & $\times$ \\
Zhao et al. \cite{zhao2025impacts,zhao2026discriminatory} & Food delivery dispatch \& payment & \checkmark & $\times$ & \checkmark & $\times$ & $\times$ & $\times$  \\
\midrule
\textbf{RideGym (ours)}                    & \textbf{MARL dispatch benchmark} & \checkmark & \checkmark & \checkmark & \checkmark & \checkmark & \checkmark \\
\bottomrule
\end{tabular}
\end{table*}

\section{Visualization Example} \label{sec:example}
\begin{figure}[htbp]
\centering
\includegraphics[width=0.35\textwidth]{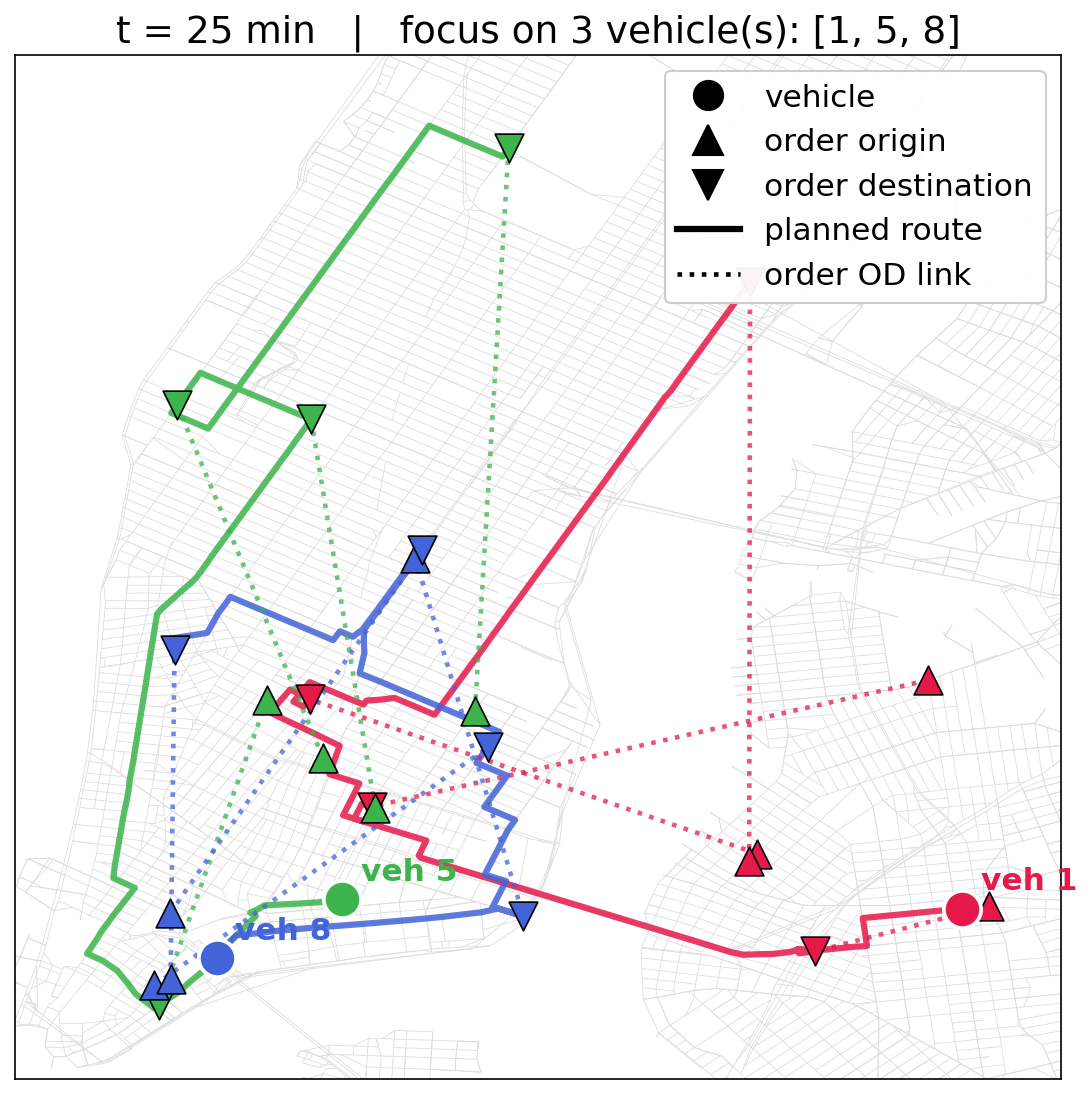}
\caption{A single-frame visualization of three focused vehicles under the KM dispatch baseline. Each vehicle is drawn in its own color, the disc marks its current location, the solid line its planned route along the real road network, and the up/down triangles the origin (\(l^o\)) and destination (\(l^d\)) of each order it is currently serving, joined by a dashed origin-destination link.}
\label{fig:visual}
\end{figure}

Fig.~\ref{fig:visual} shows a representative single frame, focusing on three vehicles served by the KM dispatch baseline. To keep the picture legible, each vehicle and its associated elements share a unique color, while the marker {shape} encodes semantics: a filled disc denotes the vehicle's current location \(l^p_{i,t}\), an up-triangle an order origin \(l^o\), and a down-triangle an order destination \(l^d\). For every vehicle, a solid line traces its planned route along the real road network, computed by the route planner of Section~\ref{sec:env-impl} under the pickup-before-drop-off constraint, so one can directly read off the sequence of stops it will visit. For each order the vehicle is currently carrying or heading to pick up, a dashed line links that order's origin and destination, making the demand each vehicle is committed to immediately apparent. Together, color (which vehicle), shape (vehicle, origin, or destination), and the two line styles (planned route vs. origin-destination link) allow a reader to quickly disentangle which vehicle is serving which orders and along what path.

\section{Experiment Details}
\subsection{Benchmark Methods Introduction} \label{sec:benchmark}
For model-based methods, we select the following baselines:
\begin{itemize}[left=0pt]
\item \textbf{Random:} The simplest baseline, which randomly assigns orders to vehicles located within the same region. In our experiments, the region is defined as a circle with a radius of 1 km.
\item \textbf{Greedy \cite{kalyanasundaram1993online}:} A fast yet effective heuristic that sequentially assigns each order to the nearest available vehicle.
\item \textbf{Kuhn–Munkres (KM) \cite{kuhn1955hungarian}:} A bipartite matching-based dispatch method that minimizes the total matching distance between vehicles and orders.
\item \textbf{Gale–Shapley (GS) \cite{gale1962college,yue2024end}:} An online matching algorithm used in Didi, where driver preferences over orders are based on order price (proportional to distance and passenger count), and order preferences over drivers are based on current distance (related to waiting time).
\end{itemize}

For MARL-based methods, we select the following order dispatch approached:
\begin{itemize}[left=0pt]
\item \textbf{REDA \cite{holder2025multi}:} A representative IDDQN-based baseline adapted for order dispatch, which employs an MLP to estimate the Q-value for each vehicle-order pair and subsequently applies bipartite matching to maximize the total Q-value under the independent learning assumption.
\item \textbf{BMG-Q \cite{hu2025bmg}:} Building on IDDQN, BMG-Q incorporates neighborhood information via a modified Graph Attention Network (GAT) \cite{velivckovic2017graph}. Following the original paper, we set the neighborhood size to 30.
\item \textbf{MF-DDQN \cite{li2019efficient}:} Similar to BMG-Q, MF-DDQN captures neighborhood behavior using a mean-field action. We follow the original implementation by setting the neighborhood size to 30 and using the average order information as the mean-field action.
\item \textbf{Assignment-Net \cite{zhao2025impacts,zhao2026discriminatory}:} A network architecture specifically designed for order dispatch, which uses an LSTM to encode sequential information and an MLP with ARL \cite{chen2024modelling} enhancement to encode non-sequential features. A QK-Attention \cite{zhao2025crossfi} module is then employed to compute the Q-value for each vehicle-order pair via a mutual attention mechanism.
\item \textbf{CV-Net \cite{tang2019deep}:} Designed for large-scale ride-hailing tasks, CV-Net introduces a cerebellar embedding module to learn informative region representations across multiple scales, in contrast to directly encoding raw coordinate values. Specifically, to align with our state space, we use three cerebellar embeddings for location encoding at different scales, and employ a linear layer to encode other information.
\end{itemize}
In our implementation, both Assignment-Net and CV-Net adopt the same training paradigm as IDDQN, with only the network architecture being modified. For these two methods, our primary goal is to examine the impact of architectural design on performance; therefore, we omit the Lipschitz normalization loss originally proposed in CV-Net, ensuring that all compared benchmarks share an identical loss function for a fair comparison.

We note that some recent SOTA methods and classical algorithms are omitted from our benchmark for the following reasons: (i) The primary contribution of this paper is the design of a standardized Gym-like environment, rather than an exhaustive benchmark evaluation. (ii) Certain classical methods, such as \cite{xu2018large}, rely on policy iteration and require frequent reward function evaluations. While this is feasible in ride-hailing settings, it becomes computationally expensive in ride-sharing scenarios, where the reward function is significantly more complex due to the joint consideration of immediate income and future detour costs. (iii) Recent SOTA methods, such as Triple-BERT \cite{zhao2026triplebert}, challenge the conventional MARL paradigm by proposing a centralized Single-Agent RL (SARL) solution. However, as the mainstream paradigm in this domain remains MARL-based, we defer the support for SARL algorithms to future work.

\subsection{Evaluation Metrics} \label{sec:metric}
We evaluate all methods using the following four metrics:
\begin{itemize}[left=0pt]
\item \textbf{Service Rate:} The proportion of orders that are successfully assigned to a vehicle, computed as the number of dispatched orders divided by the total number of requested orders. A higher service rate indicates that more passenger demand is being served.
\item \textbf{Completion Rate:} The proportion of orders that are not only assigned but also finished within the episode, computed as the number of completed trips divided by the total number of requested orders. It reflects the platform's end-to-end ability to actually fulfill demand rather than merely matching it.
\item \textbf{Wait Time:} The average time elapsed between an order being requested and the assigned vehicle arriving at the pickup location. A shorter wait time corresponds to a better passenger pickup experience.
\item \textbf{Detour Time:} The average extra travel time incurred relative to the shortest path from origin to destination, caused by pooling detours. A shorter detour time indicates more efficient trip execution and lower passenger inconvenience.
\item \textbf{Simulation Time:} The average runtime per episode, capturing the computational cost of each method.
\end{itemize}

\section{Discussions} \label{sec:discussion}
As the first open-source, standardized Gym-style simulator tailored to (MA)RL-based ride-sharing order dispatch, RideGym strives to support a wide range of functionalities—such as personalized vehicle settings and multiple passengers per order—aligning with the experimental setups of most existing works \cite{hu2025bmg,enders2023hybrid,hao2022hierarchical}. Nevertheless, several promising directions remain for future extensions. First, to enhance realism, the simulator could incorporate dynamic traffic conditions (e.g., congestion) and emergency scenarios, possibly by integrating with microscopic traffic simulators such as SUMO \cite{behrisch2011sumo}. Second, the framework can be extended to jointly address other operational tasks beyond order dispatch, including vehicle relocation, charging/fueling scheduling, dynamic decision windows, and pricing (to passengers) and payment (to drivers) mechanisms. Furthermore, multi-modal transportation options, such as taxi-subway integration \cite{hu2025coordinating} and land-air mobility \cite{liu2025joint}, could also be considered. Given the structural similarity between ride-sharing and logistics or food delivery systems, our simulator could potentially be adapted to support these tasks or even hybrid human-goods transportation scenarios \cite{liu2025piggyback}.

From the perspective of academic community development, we strongly encourage researchers to open-source their code and report detailed experimental configurations. In our benchmark study, we observe that many methods are sensitive to hyper-parameters, like the neighborhood size in BMG-Q and MF-DDQN, and the cerebellar resolution in CV-Net, making it impractical for a single research group to thoroughly reproduce and tune all such parameters across multiple baselines. Moreover, our experimental findings reveal that the choice of exploration noise can significantly affect the performance of MARL-based solutions, a factor often overlooked in prior work. This raises important questions: How can we systematically design suitable exploration noise rather than rely on heuristic choices? Is the optimal noise type consistent across different methods? More critically, different noise types can alter the relative performance ordering of methods, confounding the effect of algorithmic innovations and complicating fair comparison. These observations highlight the need for standardized evaluation protocols that mitigate the confounding influence of noise design, and we hope our work serves as a step toward more reproducible and equitable benchmarking practices.









\end{document}